\documentclass[a4paper,11pt]{article}
\usepackage{nfraconf,graphicx,cite}

\begin{document}
\baselineskip=13pt

\title{Limitations of ad~hoc ``SKA+VLBI'' configurations \& the need to extend
  SKA to trans-continental dimensions} 

\author{ M.A. Garrett}
\address{Joint Institute for VLBI in Europe,
Postbus 2, 
7990 AA Dwingeloo,
The Netherlands. \\
E-mail: garrett@jive.nfra.nl}


\maketitle

\abstract{The angular resolution of the proposed Square Kilometre
  Array, SKA, must be extended towards the milliarcsecond scale if it
  is to resolve the distant starburst galaxies that are likely to
  dominate the radio source counts at micro and sub-microJy flux
  levels. This paper considers the best way of extending SKA's angular
  resolution towards the milliarcsecond scale. Two possible SKA-VLBI
  configurations have been investigated and simulated SKA and SKA-VLBI
  visibility data sets generated. The effects of non-uniform data
  weighting on the associated images are considered. The results
  suggest that the preferred option is for SKA to be extended to
  trans-continental dimensions. By retaining 50\% of the array's
  collecting area within a region no larger than 50~km, the surface
  brightness sensitivity of the array at arcsec resolution is hardly
  compromised. In this configuration SKA's capabilities are impressive:
  in a single 12 hour run, between $ 100-1000$ sources will be
  simultaneously detected and imaged with arcsecond, sub-arcsecond and
  milliarcsecond resolution.}

\section{Introduction}

The technique of Very Long Baseline Interferometry (VLBI) permits
astronomers to generate milli and sub-milliarcsecond resolution images
of galactic and extra-galactic radio sources. VLBI has evolved rapidly
in the last decade with significant improvements in resolution,
polarisation imaging and spectral-line capabilities.  In terms of raw
sensitivity, however, the gains have been more modest, especially at
cm-wavelengths. Although the technique of phase-referencing has
recently permitted the detection and (limited) imaging of sources at
the mJy flux level, the current state-of-the-art r.m.s.  image noise
level is still limited to $\sim 30\mu$Jy/beam at cm-wavelengths (for
a typical on-source observing run of 12~hours).  The recent
introduction of the 1 Gbit/sec MkIV system and other technical
improvements promise to improve this by a factor of 3 or so, thus
reducing the r.m.s. image noise level to $\sim 10\mu$Jy/beam.

While this is all very encouraging, a more sobering thought is that
even at these r.m.s. noise levels, the overlap between the radio sky and
the sky at optical and infra-red wavebands is rather limited. It is
only by going deeper -- much deeper -- that the optical and radio
source counts become comparable. If complimentary observations are to
be achieved, and radio astronomy is to remain at the very forefront of
astrophysical research, it is imperative that noise levels are reduced
by at least two orders of magnitude.

The Square Kilometer Array, SKA, currently offers the best possibility
of achieving these kind of noise levels. However, sensitivity is not
the only issue, one must also consider what angular resolution is
required - not simply to avoid the limitations imposed by source
confusion but to properly investigate the radio morphology of the
sources that will dominate the $\mu$Jy and sub-$\mu$Jy radio source
population.  

The tendency for faint sources to be considerably smaller than their
brighter counterparts has been known for some time (Oort 1987
\cite{paper01} and Fletcher et al. 1998 \cite{paper02}), and is
strikingly confirmed by the relatively high detection rates of recent
VLBI surveys of faint mJy radio sources (Garrington, Garrett \&
Polatidis 1999 \cite{paper03}). For compact AGN this phenomenon can be
easily understood in terms of synchrotron self-absorption theory: for a
given magnetic field strength, smaller sources will also be fainter
sources. High resolution imaging is therefore of considerable
importance to the study of faint AGN.

Perhaps a more significant factor in this discussion, however, is the
emergence of a new population of radio sources as suggested by the
flattening radio source counts at sub-mJy flux density levels, now
confirmed with the recent radio studies of the Hubble Deep Field
(Richards et al. 1998 \cite{paper04} and Muxlow et al. 1999
\cite{paper05}). These pioneering observations strongly suggest that
the bulk of the $\mu$Jy radio source population is dominated by distant
starburst galaxies, rather than AGN.

\begin{figure}
  \begin{center}
    \includegraphics[scale=0.46, angle=-90]{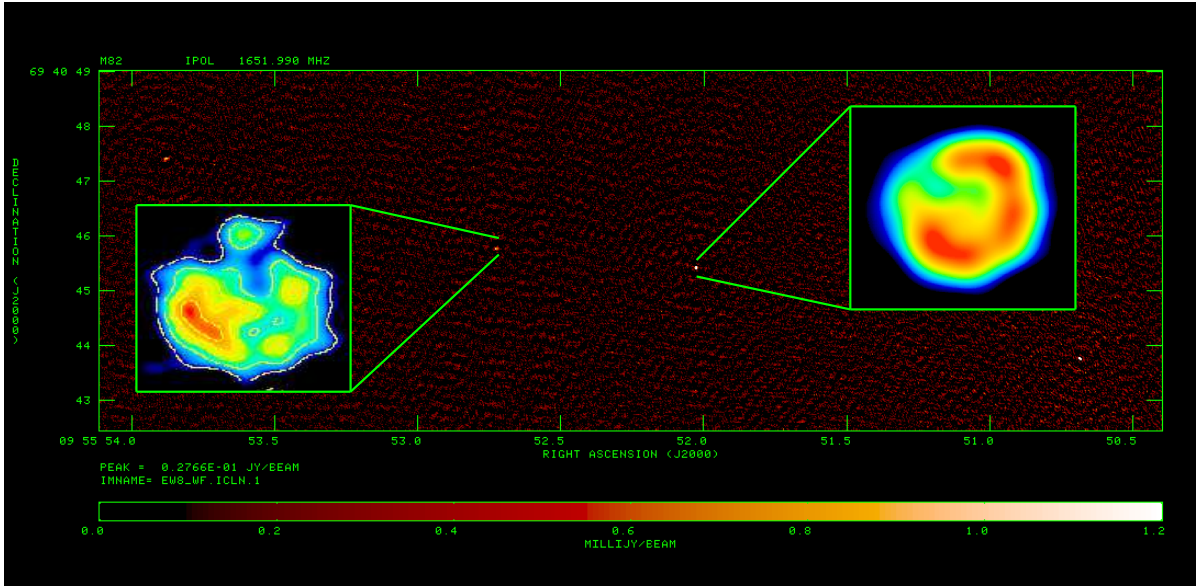}
\caption{\small Wide-field EVN $\lambda 18$~cm observations of M82 (Pedlar et
  al. 1999) reveal SNR shells ranging in diameter from 0.6 to 1.4 pc.
  Evidence for clear expansion in the remnant highlighted in the upper
  right-hand corner of this figure, is consistent with a SN event in
  the early 1960's. The image is a good example of the current VLBI
  state-of-the-art: the entire 1 arcminute field was imaged in
  wide-field, phase-reference mode. The faintest remnant has a total
  flux density of 2.6 mJy and the r.m.s. noise in the image is $\sim
  50\mu$Jy/beam. }
    \label{fig1}
  \end{center}
\end{figure}

The nearest and best studied starburst galaxy is undoubtedly M82.
Located only $\sim 3$~Mpc away, its radio emission is concentrated
within the central few kpc of the galaxy \cite{paper06} and is
dominated by radio emission from both recent and relic supernova
remnants (SNRs).  Fig.~\ref{fig1} shows a superb, wide-field EVN
$\lambda 18$~cm image of M82 produced by Pedlar et al. (1999)
\cite{paper07}.

If M82 is typical of higher redshift starbursts (and recent VLBI
observations by Smith et al. 1999 \cite{paper08} of the more vigorous
and distant starburst, Arp 220, suggest that it may well be), we can
expect to detect with SKA individual SNR in starburst galaxies out to
cosmological distances - at least in terms of sensitivity. But this is
only part of the story. If the SNR are distributed on scales similar to
that observed in both Arp 220 and M82, then at $z = 1.5$ the bulk of
the radio emission will occupy a region of sky no greater than 60
milliarcseconds across. Thus in order to properly resolve these systems
into their constituent parts, resolutions of a few mas are required.

While this is not the only argument for high resolution SKA
observations (see contributions by Gurvits, Krichbaum, Snellen,
Phinney, Roy, Koopmans \& Fender - these proceedings) it is a very
powerful one: the idea that SKA will only barely resolve the dominant
sources of radio emission in the sky (Wilkinson's ``bread and butter
sources'' - these proceedings) is surely unthinkable, at least for a
true ``next-generation'' instrument.

In this paper, I discuss the ways in which the angular resolution of
SKA can be extended towards the milliarcsecond scale. I do not make the
conventional assumption that SKA's only contribution to the field of
VLBI is as an ultra-sensitive, phased-array ``add-on'' to existing VLBI
networks. Although this scenario is considered other options are
investigated and indeed preferred, including the extension of SKA to
trans-continental dimensions.

I first consider in section 2 a few minor technicalities regarding SKA
and VLBI baseline sensitivity (in particular the possibility of
employing in-beam phase referencing techniques), and the need for a
wide-field approach to VLBI observations at these sub-$\mu$Jy levels.
In Section 3 I present the first realistic simulations of various
SKA-VLBI configurations including an extended version of SKA (designated
``SKA$^{++}$''), in which half of the SKA antennas are located within
an array of 50~km and the other half are separated by trans-continental
distances.  A discussion of the main results and conclusions are
presented in sections 4 and 5 respectively.

\section{SKA-VLBI: minor technicalities} 

Throughout this paper I adopt the nominal SKA parameters of Taylor \& Braun
(1999) \cite{paper09} {\it i.e.} thirty, 200-m diameter elements with a
total observing bandwidth of 1500~MHz at $\lambda 6$~cm, 2-bit sampled
data and a total sensitivity figure of $2\times10^{4}$m$^{2}$/K.  In
this section we discuss some minor technicalities that have not been
previously considered in earlier SKA-VLBI discussions.

\subsection{SKA \& VLBI baseline sensitivity} 

The $7\sigma$ baseline detection level between a phased-array SKA,
({\it i.e.} SKA$_{\rm PA}$ with 80\% of the total collecting area
formed by those antenna elements lying within 50~km of each other, SEFD
$\sim 0.17$~Jy) and a single 25-m VLBA antenna (SEFD $\sim 290$Jy)
is $\sim 60\mu$Jy, assuming a coherent
integration time of 300 seconds at $\lambda 6$~cm. At these levels of
sensitivity the radio source counts are fairly well understood: the
recent VLA HDF observations of Richards et al. (1998) \cite{paper04},
together with earlier VLA observations including those of Windhorst et
al. (1995) \cite{paper10}, suggest the source count is of order
20S($\mu$Jy)$^{-1}$~arcmin$^{-2}$.  Thus within the {\sc FWHM} of a 25-m
antenna's primary beam, we can expect to find at least 15 sources above
the $10\sigma$ noise level, of which 5 can reasonably be expected to be
stronger than the $30\sigma$ noise level.  Naturally these latter
sources can be used as ``in-beam'' (phase) calibrators, able to provide
continuous and accurate instrumental corrections without the need for
conventional phase-referencing (note that in this scenario the multiple
beam capability of SKA in its phased-up mode is assumed since the field
of view of the phased-array is otherwise rather limited). At
$\lambda$18cm the situation is even better with at least 45 potential
calibrator sources in the primary beam of a 25-m antenna. In short,
in-beam calibration will be possible in the vast majority of cases at
cm wavelengths.

\subsection{Imaging large fields-of-view} 

At the SKA detection level of 100~nanoJy we can (with a little
extrapolation!)  predict a source count of at least $\sim 100$
sources/arcmin$^{2}$.  Clearly the sky will be densely populated with
radio sources separated by only a few arcseconds, perhaps less, if
clustering is important (as one might expect). Independent of whether
we consider SKA as a standalone array (with baseline lengths of at
least 1000 km) or as part of a VLBI network, we can expect to image
hundreds of sources simultaneously from just a single area of sky
covered by one (single element) SKA beam.  Already the application of
wide-field imaging techniques is beginning to find a place in VLBI
(Garrett et al. 1999 \cite{paper11}) but in the era of SKA, a
wide-field imaging mode will be the {\it de facto} mode of operation -
even at milliarcsecond resolutions. This will require {\it at least}
the full spectral resolution of SKA at the longest wavelengths
($10^{4}$ spectral channels) and sub-second integration times at the
shortest cm wavelengths in order to avoid smearing. A typical 12 hour run
by the SKA-VLBI configurations discussed further in this paper, will
result in a substantial but hopefully not unmanageable data size of
$\sim 1$~Tera Byte.

\section{SKA \& VLBI: issues of sensitivity and weight}

Various authors have previously considered the sensitivity gain one
achieves by including SKA as part of a large VLBI network or extending
it to trans-continental baselines ({\it e.g.} Schilizzi \& Gurvits,
section 2.5.2 in Taylor and Braun \cite{paper09}). In this section we
extend these calculations by taking into account the effect of data
weighting and the necessary trade-off between sensitivity and
resolution for some proposed SKA-VLBI configurations.

\subsection{Weighty matters} 

At face value the inclusion of SKA as part of a large VLBI network
results in an array with superb uv-coverage, high resolution and
sub-$\mu$Jy sensitivity. However, predictions of image noise levels
(and uv-coverage) which do not take into account the relative weights
of the contributing baselines, can be misleading.  For example, if we
consider an array formed by the individual SKA elements (SKA$_{1}$), in
the nominal configuration of Taylor \& Braun, observing together with a
global VLBI array (GVLBI), then for naturally weighted data the array
is entirely dominated by the very sensitive baselines formed between
SKA elements. Since the vast majority of these present baseline lengths
of order 50~km or less, the dirty beam associated with such naturally
weighted data does not even begin to provide the sort of milliarcsecond
(mas) resolution expected from a VLBI array of global dimensions.
Alternatively if SKA is included as a phased-array, SKA$_{\rm PA}$, the
situation is even more extreme in terms of the effective uv-coverage
since only the SKA$_{\rm PA}$ baselines actually contribute to the
synthesised image.  In either case, the only way to achieve uniform
uv-coverage is to abandon natural weighting and re-weight ({\it i.e.}
weight-up) the noisier baselines, thus increasing the image noise level
by factors of several - well beyond the original expectation.

\subsection{SKA-VLBI data simulations} 

In order to investigate these effects semi-quantitatively, I have
generated three simulated SKA (and VLBI) visibility data sets.  In
order to serve as a reference point, simulated data were first
generated for the nominal SKA configuration of Taylor \& Braun (1999)
\cite{paper09}.  Two additional options were considered with respect to how
SKA provides high sensitivity observations with milliarcsecond scale
resolution: (i) SKA contributes as a sensitive, phased-array
``add-on'' to the existing global VLBI network (``SKA$_{\rm
  PA}$-GVLBI''), and (ii) SKA is extended to trans-continental
baselines, SKA$^{++}$ but 1/2 of the antennas still remain within 50~km
of each other in order to maintain good brightness sensitivity at
arcsecond resolution.
 
\subsubsection{Source Model} 

\begin{figure}
  \begin{center}
    \includegraphics[scale=0.5, angle=-90]{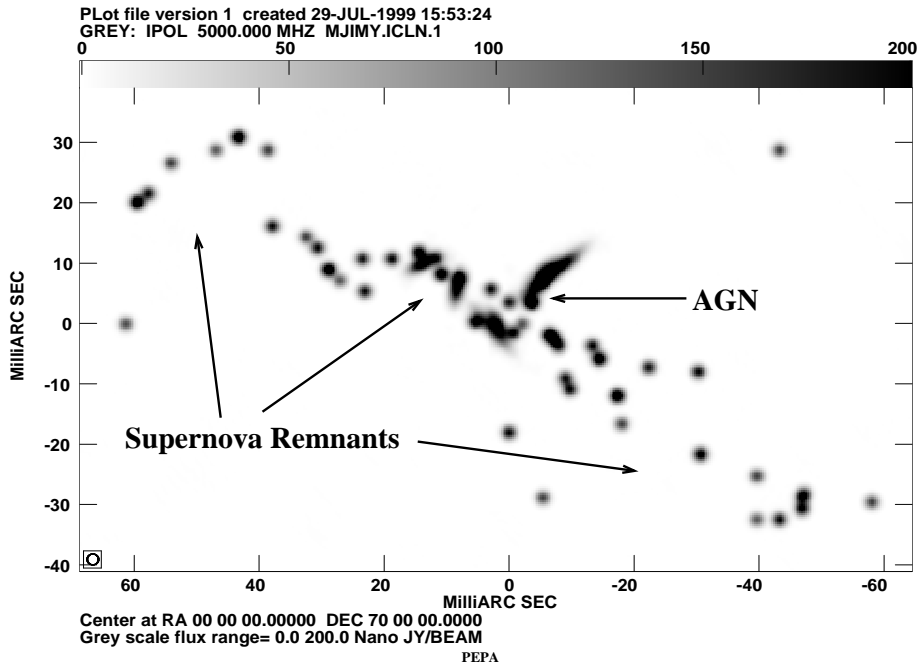}
    \caption{\small The model source convolved with a 1.8 mas circular 
beam. The source spans a region 120 milliarcseconds across. The total 
flux is $14\mu$Jy.}
    \label{fig2}
  \end{center}
\end{figure}

The source model used to produce the simulated $\lambda 6$~cm data
is shown in Fig.~\ref{fig2}. It 
represents a ``best guess'' of what the $\mu$Jy source population might
look like -- essentially it is based on an M82/Arp~220-type starburst,
projected back to a redshift of $z \sim 1.5-2$. The radio emission is
concentrated within the inner few kpc (120 mas) of the galaxy, and
dominated by young SNRs, relic SNRs and compact HII regions. In
addition, I have added a low-luminosity AGN slightly offset from the
plane of star formation which accounts for $\sim 20\%$ of the total
flux density of 14~$\mu$Jy.  Although an AGN has not yet been
identified in either Arp 220 or M82, it might not be too surprising to
find such low-luminosity AGN in some starburst galaxies. The faint
radio sources identified with starburst galaxies in the HDF, appear to
have very distorted morphologies, suggesting they were recently
involved in interactions with their nearest companions or complete
mergers.  While this is certainly a trigger for rapid bursts of star
formation, it may also initiate AGN activity (or re-initiate it) in the
centres of these galaxies.

\subsubsection{Data Generation} 

The AIPS task {\sc UVCON} was used to generate the $\lambda 6$~cm
simulated data sets.  The nominal SKA configuration (Taylor \& Braun
1999 \cite{paper09}) was initialy assumed, with the array arbitrarily
centred on Dwingeloo, the Netherlands. Data were generated between hour
angles of $\pm6$ hours (as calculated at the centre of the array). {\sc
  UVCON} adds Gaussian noise to each visibility based on the specified
antenna characteristics (diameters, efficiency, noise temperature, data
sampling/rate etc). For the elements of SKA the following parameters
were chosen: 30 identical elements of 200~m diameter and 60~K system
temperatures with a combined sensitivity figure of $\sim
2\times10^{4}$m$^{2}$/K.

\subsubsection{Simulated SKA Images} 

\begin{figure}
  \begin{center}
    \includegraphics[scale=0.5, angle=-90]{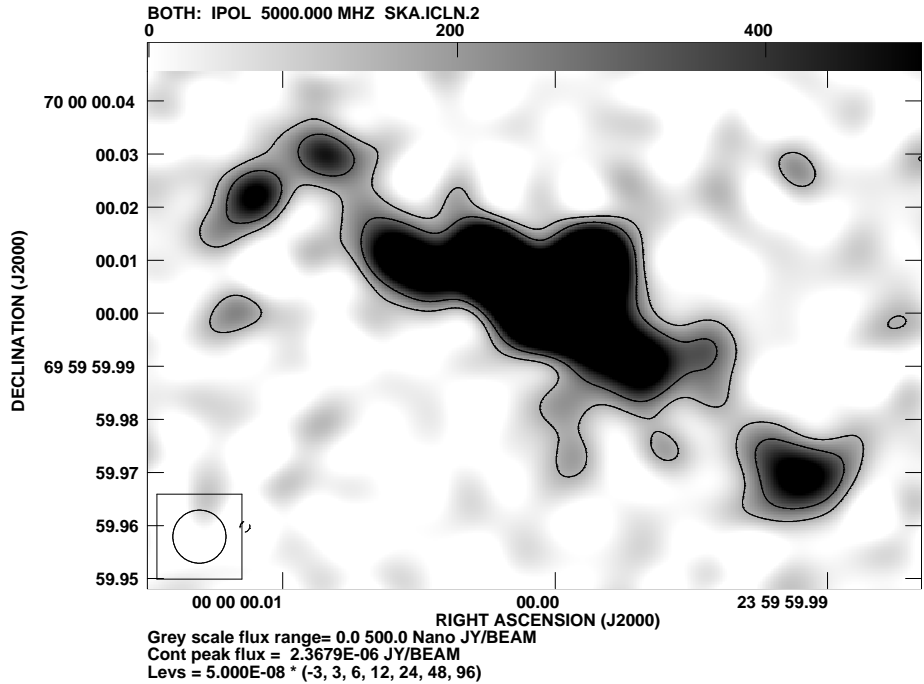}
    \caption{\small A uniformly weighted 
      $\lambda 6$~cm image produced by the nominal SKA configuration of
      the starburst model source. The resolution is $\sim 10$~mas but 
      does not adequately resolve the source. The r.m.s. noise level 
      is $\sim 0.05\mu$Jy/beam.}
    \label{fig3}
  \end{center}
\end{figure}

Fig.~\ref{fig3} shows a simulated $\lambda 6$~cm image of the model
source generated by the nominal SKA configuration. The data were
Fourier transformed and CLEANed using the AIPS task {\sc IMAGR}.  The
image was produced with (Robust=-2) uniform weighting (see Briggs 1995
\cite{paper12} for a discussion of Robust weighting).  This weights the
data at a level which is intermediate between natural weighting (the
case in which visibility weights are simply proportion to the inverse
of the r.m.s. noise squared) and pure uniform weighting (all data
points have equal weights irrespective of their variance and the local
data density in the uv-plane). This weighting is necessary in order for
the nominal SKA configuration to provide the 10~mas resolution one
expects for an array in which the longest baselines are $\sim 1000$~km.
The naturally weighted image provides only 20~mas resolution since the
uv-plane is so densely populated by the inner 50~km region of the array
where 80\% of the collecting area resides. However, even the 10~mas
resolution obtained from the uniformly weighted data is not sufficient
to do much better than partially resolve the radio source.  The noise
in this image is $\sim 0.05 \mu$Jy/beam, almost twice as high as the
noise in the naturally weighted image.

\subsubsection{Simulated ``SKA$_{\rm PA}$+GVLBI'' Images} 

\begin{figure}
  \begin{center}
    \includegraphics[scale=0.5, angle=-90]{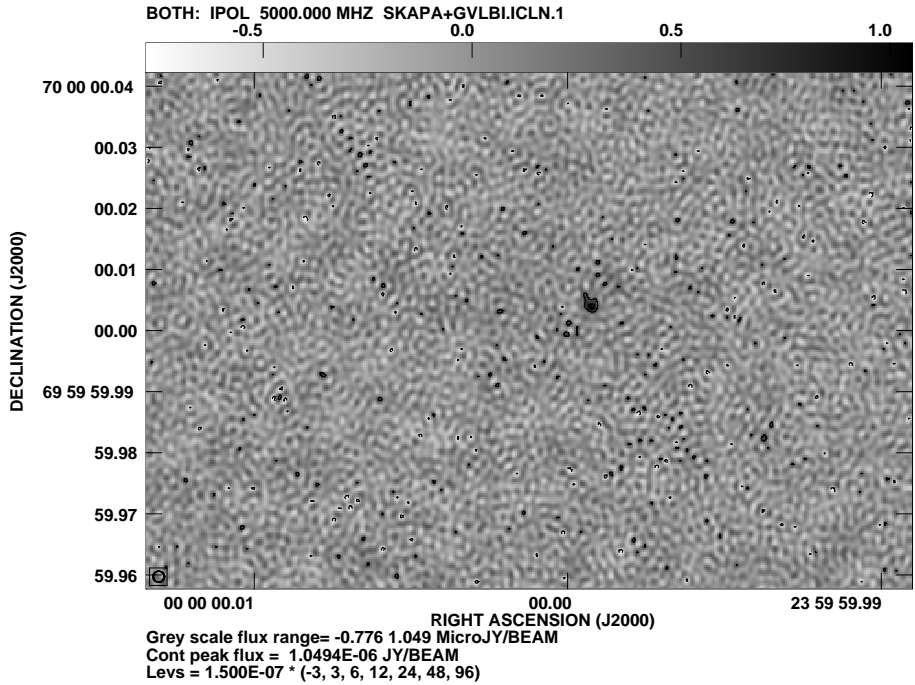}
    \caption{\small The SKA$_{\rm PA}$+GVLBI uniformly weighted 
      $\lambda 6$~cm image of the Starburst model radio source
      described in the text.  The resolution is $\sim 1$~mas but the
      combined array is only able to detect the core of the AGN. The
      fainter SNRs and the extended AGN jet go undetected. The r.m.s.
      noise level is $\sim 0.17\mu$Jy/beam.}
    \label{fig4}
  \end{center}
\end{figure}

Fig.~\ref{fig4} shows a simulated $\lambda 6$~cm image of the model
source generated by a global VLBI network supplemented by the inner
80\% of the nominal SKA configuration, phased-up to form a single,
highly sensitive VLBI antenna, ``SKA$_{\rm PA}$''. This is the
traditional SKA-VLBI configuration that is often assumed to be SKA's
default contribution to VLBI.  The global VLBI network used in these
simulations includes 17 of the largest antennas in the world, including
the Effelsberg 100-m, VLA$_{27}$, Greenbank 100-m, DSN 70-m and the new
70-m and 45-m antennas currently under construction in Sardinia (IRA)
and Yebes (OAN). As for the previous SKA simulation, we assume the VLBI
antennas can also deliver or record data at 6~Gbits/sec (a
2-bit/4-level sampled, single polarisation, 1500~MHz wide IF band). The
naturally weighted image has an r.m.s. noise level of 0.17$\mu$Jy/beam
and provides a resolution of 1~mas. The core of the AGN is barely
detected but the other sources fall well below the noise level. The
image (and the effective uv-coverage) are completely dominated by
SKA$_{\rm PA}$ baselines, the other inter-VLBI antenna baselines have
no effect on the image whatsoever.  Although this latter effect is yet
to be investigated in any detail, the ability of this array to image
even moderately extended structures is likely to be limited.  Brute
force modification of the antenna weights would be required in order to
improve the effective coverage but the corresponding impact on
sensitivity would be severe.

\begin{figure}
  \begin{center}
    \includegraphics[scale=0.5, angle=-90]{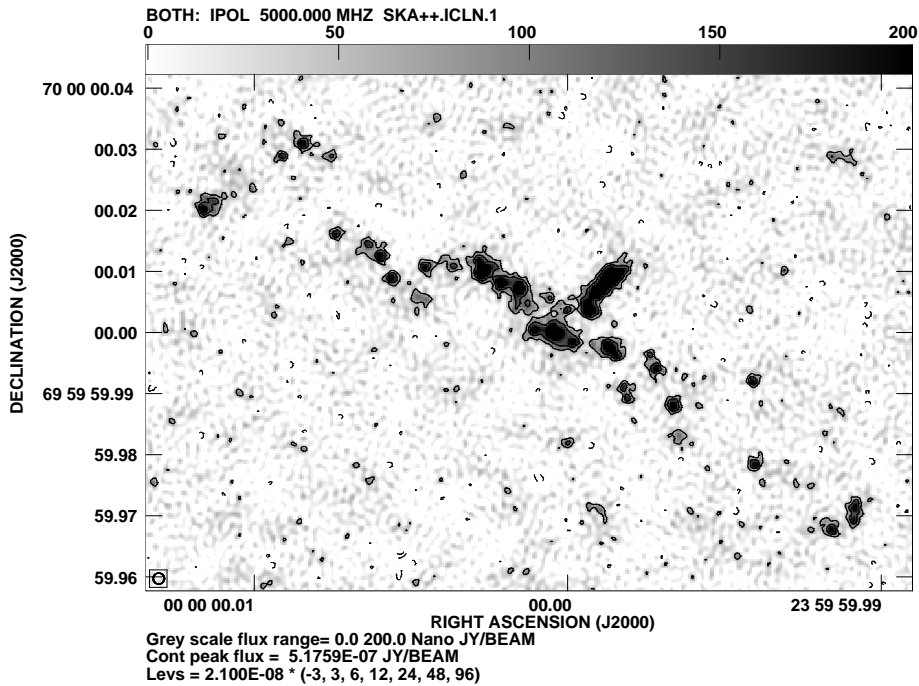}
    \caption{\small The SKA$^{++}$ uniformly weighted $\lambda 6$~cm
      image of the Starburst model radio source described in the text.
      The resolution is $\sim 1.8$~mas permitting the structure source
      to be revealed in some detail. The r.m.s. noise level is $\sim
      0.02\mu$Jy/beam.}
    \label{fig5}
  \end{center}
\end{figure}

\subsubsection{Simulated ``SKA$^{++}$'' Images} 

Fig.~\ref{fig5} shows a simulated $\lambda 6$~cm image of the model
source generated by an extended SKA configuration, ``SKA$^{++}$''. In
this scenario half of the SKA antennas remain within the inner 50~km of
the array but the other half are distributed around the world (in this
case the precise locations are a subset of the current EVN and VLBA
antenna sites).  Note that the inner elements of SKA contribute to the
observations as {\it individual} antennas, not as a phased array
(though note that in order to provide the in-beam calibration described
in section 2.1, an additional phased array beam may also be required).
A ``robust=0'' uniformly weighted image has an r.m.s.  noise level of
$\sim 0.02\mu$Jy/beam and provides a resolution of 1.8~mas. The
combination of high sensitivity and resolution allows us to resolve the
individual SNR from each other and the AGN (which also shows a
two-sided jet). The vast majority of the individual SNR themselves
remain unresolved; space VLBI resolutions such as those achievable by
the proposed ARISE mission (Ulvestad \& Linfield 1998 \cite{paper13})
might be able to detect and thus resolve the brighter remnants (the
sensitivity of a combined SKA+ARISE configuration requires its own
detailed study, see Gurvits these proceedings).


\section{Discussion} 

The SKA can make a useful contribution to high resolution radio
astronomy. However, the nominal SKA configuration with baseline lengths
$< 1000$~km may not provide enough resolution to adequately resolve
the structure of the vast majority of the faint, extragalactic radio
sources it detects.  In my opinion this is a major flaw in the proposed
configuration. Higher resolution can be achieved in at least two ways, the
conventional option is for SKA to participate within a VLBI network as
a highly sensitive phased-array add-on. The second less conventional,
but in my opinion preferred option, is for SKA to be extended to
trans-continental baselines, SKA$^{++}$.

The conventional option will allow images to be made with noise levels
of $\sim 0.17\mu$Jy/beam, a factor of 60 better than what can be
achieved by VLBI today, even in the era of 1~Gbit/sec MkIV recording. The
uv-coverage will, however, remain limited, and there is a danger that
the coordination and flexibility of the network will be plagued by the
problems that beset existing, non-homogeneous, ad~hoc arrays.

The SKA$^{++}$ option will allow images to be made with noise levels
around $ 0.02\mu$Jy/beam, a factor of 500 better than what can be
achieved today and almost an order of magnitude better than 
the conventional phased-array option. In the SKA$^{++}$
configuration described here, half the antennas are still located within
the inner 50~km region of the array, thus satisfying other SKA
programmes which require high surface brightness sensitivity at
arcsecond resolution. This homogeneous array offers superb uv-coverage,
flexibility in operation and all the other benefits associated with
SKA, in particular, multiple beams for phase-referencing (although at
wavelengths $\geq 6$~cm this may not be required since there will
almost always be enough ``in-beam'' calibrators). The possibilities
arising from a SKA$^{++}$ instrument are quite simply staggering: with
a 1 arcmin field of view, SKA$^{++}$ in a single 12 hour run, could
easily detect and image over $\sim 100-1000$ sources simultaneously
with arcsecond, sub-arcsecond and milliarcsecond resolution. 

The feasibility of connecting together SKA elements in real-time over
large distances appears feasible, even by today's standards.
Considerable activity in the connection of telescopes by optical fibres
is on-going around the world with the recent link between the VLBA
antenna at Pie Town and the VLA (a distance of $\sim 100$~km), being
the most recent success story. The main difficulties are now considered
to be economic rather than technical (Whitney et al. 1999
\cite{paper14}). With the reasonable expectation that trans-continental
fibre connections will fall in price over the next 2 decades,
SKA$^{++}$ is a realistic proposal which requires serious consideration
and more detailed investigation.

\section{Summary: the need for a higher resolution SKA} 

The quest for higher angular resolution has been one of the key driving
forces in observational astronomy, together with improved sensitivity
and new spectral bands.  Despite the fact that optical telescopes have
always enjoyed a natural advantage in terms of source number counts,
the ability of radio interferometers such as the VLA, MERLIN and VLBI
to generate sub-arcsecond and milliarcsecond resolution images has
allowed them to stay at the very forefront of astrophysics.  Comparable
radio instruments, in terms of sensitivity but with inferior resolution,
have been significantly disadvantaged.

Optical astronomers are now designing the next generation of ground and
space based telescopes ({\it e.g.} the VLTI \& NGST). These will have
comparable or {\it better} resolution than that currently proposed for
SKA. Similarly, it is now clear that optical and infra-red
interferometry will take a giant leap forward in terms of sensitivity
and resolution, in the form of the armada of space-based interferometry
missions ({\it e.g.} Gaia, Darwin, SIM etc) currently proposed. 
On the same time scales envisaged for the completion of SKA, 
these next generation instruments will provide optical and
infra-red astronomers with the ability to perform micro-arcsecond
astrometry (allowing the direct detection of nearby extra-solar
planets) and sub-milliarsecond resolution imaging of a wide variety of
celestial objects. The importance of complimentary, high resolution
radio observations will become clear as the surfaces of
nearby stars, the ejecta of novae and supernovae, accretion disks and
jets around young stars and x-ray binary systems, not to mention the
environment around the central engines of extra-galactic objects
(normal galaxies and AGN) become the favoured targets of these space-based 
instruments. 

The next generation of radio telescope will surely provide astronomers
with unprecedented sensitivity - that much is clear. However, the
majority of radio sources it detects will most likely require
milliarcsecond resolution to be adequately resolved. Simply relying on
occasional, ad~hoc ``SKA+VLBI'' observations to provide this resolution
is not, in my opinion, a satisfactory solution.  A self-contained SKA
can provide milliarcsecond resolution by extending the array to
trans-continental dimensions. By retaining 50\% of the array's
collecting area within a region no larger than 50~km, the surface
brightness sensitivity of the array at arcsec resolution is hardly
compromised. In this way SKA can be a truly global, next generation
radio telescope with unrivaled capabilities over a wide range of
angular resolution and surface brightness sensitivity.

\section*{Acknowledgments}

I'd like to thank Richard Schilizzi for reading the text of this paper
critically, and for suggesting several useful improvements that were
incorporated into the final version.


\section*{References}

\begin{thebibliography}{99}

\bibitem{paper01}Oort, M.,  1988, A\&A, 193, 50.  

\bibitem{paper02}Fletcher, A., Burke, B.,
  Conner, S. et al.  in: Bremer, M., Jackson, N., \& Perez-Fournon, I., 1998,
  (eds) ``Observational Cosmology with the New Radio Surveys'', Kluwer,
  Dordrecht, p. 255.  

\bibitem{paper03}Garrington, S.T., Garrett, M.A. \& Polatidis, A.G.,
  1999, NewAR, in press (astro-ph/9906158) 

\bibitem{paper04}Richards, E.A., Kellermann,
  K.I., Fomalont, E.B., Windhorst, R.A., \& Partridge, R.B. 1998, AJ,
  116, 1039.  
\bibitem{paper05} Muxlow, T.W.B., Wilkinson, P.N.,
  Richards, A.M.S., Kellerman, K.I, Richards, E.A., \& Garrett, M.A.,
  1999, NewAR, in press 

\bibitem{paper06} Muxlow, T.W.B., Pedlar, A.,
  Wilkinson, P.N., Axon, D.J., Sanders, E.M., \& de Bruyn, A.G., 1994,
  MNRAS, 266, 455.
  
\bibitem{paper07}Pedlar, A., Muxlow, T.W.B., Garrett, M.A., Diamond,
  P., Wills, K.A., Wilkinson, P.N., \& Alef, W., 1999, MNRAS, in press.

\bibitem{paper08} 
Smith, H.E., Lonsdale, C.J., Lonsdale, C.J. \& Diamond, P.J., 1998, 
ApJ, 493, L17. 

\bibitem{paper09}Taylor, A.R. \& Braun, R., 1999, ``Science with the 
Square Km Array'' (URL: www.nfra.nl/conf/sciencecase.htm). 

\bibitem{paper10}Windhorst, R.A. et al., 1995, Nature, 375, 471. 
  
\bibitem{paper11} Garrett, M.A., Porcas, R.W., Pedlar, A., Muxlow,
  T.W.B., \& Garrington, S.T., 1999, NewAR, in press (astro-ph/9906108)

\bibitem{paper12} Briggs, D. 1995, PhD thesis, 
``High Fidelity Deconvolution of Moderately Resolved
Sources'',  , New Mexico Institute of Mining and Technology.
(URL: www.aoc.nrao.edu/ftp/dissertations/dbriggs/diss.html)

\bibitem{paper13} 
Ulvestad, J.S. \& Linfield, R.P., 1998, in ASP Conf. Ser. 144, 
IAU Colloq. 164: ``Radio Emission from Galactic and Extragalactic
Compact Sources'' eds. Zensus, Taylor and Wrobel, p397. 

\bibitem{paper14} Whitney, A.R., 1999, NewAR, in press.

\end{thebibliography}
\end{document}